\documentclass[fleqn,twoside]{article}
\usepackage{espcrc2}

\usepackage{graphicx}

\newcommand{\newc}{\newcommand}

\newc{\be}{\begin{equation}}
\newc{\ee}{\end{equation}}
\newc{\ba}{\begin{eqnarray}}
\newc{\ea}{\end{eqnarray}}
\newc{\ie}{{\it i.e. }}
\newc{\eg}{{\it eg }}
\newc{\etc}{{\it etc.}}
\newc{\etal}{{\it et al.}}

\newc{\ra}{\rightarrow}
\newc{\lra}{\leftrightarrow}
\newc{\no}{Nielsen-Olesen }
\newc{\lsim}{\buildrel{<}\over{\sim}}
\newc{\gsim}{\buildrel{>}\over{\sim}}

\newcommand{\AmS}{{\protect\the\textfont2
  A\kern-.1667em\lower.5ex\hbox{M}\kern-.125emS}}

\title{The Rise and Fall of the Cosmic String Theory for Cosmological Perturbations}

\author{L. Perivolaropoulos\address[UOI]{Division of Theoretical Physics,
        Department of Physics, \\
        University of Ioannina, 451 10 Ioannina, Greece}%
        \thanks{e-mail address: leandros@cc.uoi.gr \newline
      URL: http://leandros.physics.uoi.gr}}

\begin{document}

\begin{abstract}
The cosmic string theory for cosmological fluctuations is a good
example of healthy scientific progress in cosmology. It is a well
defined physically motivated model that has been tested by
cosmological observations and has been ruled out as a primary
source of primordial fluctuations. Until about fifteen years ago,
the cosmic string theory of cosmological perturbations provided
one of the two physically motivated candidate theories for the
generation of primordial perturbations. The cosmological data that
appeared during the last decade have been compared with the well
defined predictions of the theory and have ruled out cosmic
strings as a primary source of primordial cosmological
perturbations. Since cosmic strings are predicted to form after
inflation in a wide range of microphysical theories including
(supersymmetric and fundamental string theories) their
observational bounds may serve a source of serious constraints for
these theories. This is a pedagogical review of the historical
development, the main predictions of the cosmic string theory and
the constraints that have been imposed on it by cosmological
observations. Recent lensing events that could be attributed to
lighter cosmic strings are also discussed.

\vspace{1pc}
\end{abstract}

\maketitle

\section{INTRODUCTION: THE COSMIC STRING THEORY}
Cosmic strings (for a recent review see \cite{Kibble:2004hq} and
\cite{shel-vil-book,Hindmarsh:1994re,Gibbons:1990gp,Gangui:2001wc,Perivolaropoulos:1994hd,Vilenkin:1984ib,Brandenberger:1990ze}
for earlier reviews) are linear topological defects
(concentrations of energy which are stable for topological
reasons) and may have formed in the early universe during symmetry
breaking phase transitions predicted by microphysical theories (eg
Grand Unified Theories (GUTs)). In contrast to other topological
defects (walls and gauged monopoles) cosmic strings are consistent
with Big Bang cosmology and do not overclose the universe. Their
evolution generically includes a mechanism which transfers energy
from the string network to gravitational radiation through the
production and oscillation of massive string loops. The
cosmological effects of cosmic strings are completely determined
by the single free parameter of the model which is the mass per
unit length $\mu$ of the string or equivalently the dimensionless
product $G\mu$ where $G$ is Newton's constant. Due to Lorentz
invariance under boosts along the direction of the string this is
equal to the tension of the strings and is determined by the scale
of symmetry breaking that gave rise to the string network.

\subsection{Formation-Evolution of Strings}
The simplest model in which strings involving gauge fields can
form\cite{Nielsen:1973cs} is the Abelian Higgs model involving the
breaking of a $U(1)$ gauge symmetry. The Lagrangian of the Abelian
Higgs model is of the form \be {\cal L}= {1\over 2} D_\mu \Phi
D^\mu \Phi - V(\Phi)+ {1\over 4} F_{\mu \nu} F^{\mu\nu}
\label{strlang} \ee where $F_{\mu \nu} = \partial_\mu A_\nu -
\partial_\nu A_\mu$, $D_\mu \equiv
\partial_\mu - i e A_\mu$ and \be V(\Phi)= {\lambda \over 4} (|\Phi|^2 -
\eta^2)^2 \label{strpot} \ee  with $\Phi = \Phi_1 + i \Phi_2$. The
topological stability of the string (vortex in two space
dimensions) is due to the conservation of the topological
invariant (topological charge or winding number) \be
m=\int_0^{2\pi} {{d\alpha}\over {d \theta}} d\theta \ee where
$\alpha$ is an angular variable determining the orientation of
$\Phi$ ($\Phi = f\; e^{i\alpha}$) in the vacuum manifold which is
$S^1$ (circle) in the Abelian Higgs model and $\theta$ is the
azimouthal angle around the vortex in physical space.

The field configuration of a vortex may be described by the
following ansatz: \ba
\Phi &=& f(r) e^{i m \theta}\\
{\vec A} &=& {{v(r)}\over r} {\hat e}_\theta \ea with
$f(0)=v(0)=0$ (for single-valuedness of the field at $r=0$),
$f(\infty)=\eta$ and $v(\infty)=m/e$ (for finite energy). The
forms of $f(r)$ and $v(r)$ can be obtained
numerically\cite{Nielsen:1973cs} from the field equations with the
above ansatz. The energy of the vortex configuration (energy of
string per unit length) is \be \mu =
 \int d^2 x (f'^2 +{{v'^2}\over {r^2}} +
{{(ev-m)^2}\over {r^2}}f^2 +{\lambda \over 4} (f^2 -\eta^2)^2 )
\ee Clearly, $v\neq 0$ is required for finite energy \ie the gauge
field is needed in order to screen the logarithmically divergent
energy coming from the angular gradient of the scalar field. A
vortex with no gauge fields is known as {\it a global vortex} and
its energy per unit length diverges logarithmically with distance
from the string core. This divergence however is not necessarily a
problem in systems where there is a built-in scale cutoff like
systems involving interacting vortices. In such systems the
intervortex distance provides a natural cutoff scale for the
energy integral.

Even though no analytic solution has been found for the functions
$f(r)$ and $v(r)$, it is straightforward to obtain the asymptotic
form of these functions from the field equations (Nielsen-Olesen
equations). The obtained asymptotic form for $r\rightarrow \infty$
is\cite{Nielsen:1973cs} (but see Ref.
\cite{Perivolaropoulos:1993uj} for a correction to the standard
result for $\lambda/(2e)^2 > 1$) \ba
f(r) &\rightarrow & \eta - {{c_f}\over \sqrt{r}} e^{-\sqrt{\lambda} \eta r}\\
v(r) &\rightarrow &  {m\over e} -c_v \sqrt{r} e^{-e \eta r} \ea
where $c_v$ and $c_f$ are constants. Therefore the width of the
vortex is \be w\sim \eta^{-1} \ee where $\eta$ is the parameter of
the symmetry breaking potential also known as {\it the scale of
symmetry breaking}. The energy per unit length $\mu$ may also be
approximated in terms of $\eta$ as \be \mu \simeq \int_w d^2 x
V(0) \sim \eta^2 \ee The typical symmetry breaking scale for GUTs
is $\eta \simeq 10^{16} GeV$ which leads to extremely thin and
massive strings with $w \simeq  10^{-30} cm$, $\mu \simeq 10^{14}
tons/mm $ and $G\mu\simeq 10^{-6}$.

The formation of the string network occurs during the symmetry
breaking phase transition when the thermal fluctuations of the
complex field $\Phi$ drop and the field relaxes to the minimum of
its potential (vacuum manifold). The  phase of the field after its
relaxation will be randomly determined and will vary at causally
disconnected regions of the universe. After the field relaxes to
its vacuum, there will be (by causality) regions of the universe
where the field $\Phi$ will span the whole vacuum manifold as we
travel around a circle in physical space. By continuity of $\Phi$
there will be a point inside this circle where $\Phi=0$. Such a
point (and its neighbourhood) will be associated with high energy
density due to the local maximum of the potential (\ref{strpot})
at $\Phi=0$. This process has been verified by numerically
simulating\cite{Melfo:1994cv,Ye:1989dr,Ye:1990na} the evolution of
gauged and scalar fields during the phase transition in an
expanding background. By extending this argument to three
dimensions, the point becomes a line of trapped energy density
known as the {\it cosmic string}. This formation mechanism is
known as the Kibble mechanism\cite{Kibble:1976sj}.

In general cosmic strings form in field theories where there are
closed loops in the vacuum manifold $M$ that can not be shrank to
a point without leaving $M$. In homotopy theory terms, we require
that the first homotopy group of the vacuum should be non-trivial
\ie $\pi_1 (M) \neq 1$. It may be shown that this is equivalent to
$\pi_0 (H) \neq 1$ where $H$ is the unbroken group in the symmetry
breaking $G\rightarrow H$ (M=G/H).

The exact analytic treatment of the evolution of a network of
strings would involve the analytic solution of the time dependent
nonlinear field equations derived from the Lagrangian of Eq.
(\ref{strlang})  with arbitrary initial conditions. Unfortunately
no analytic solution is known for these equations even for the
simplest nontrivial ansatz of the static Nielsen-Olesen vortex.
Even the numerical solution of these equations is practically
impossible for systems of cosmological scales and with complicated
initial conditions. The obvious alternative is to resort to
realistic approximations in the numerical solution of the field
equations in cosmological systems. First, the correct initial
conditions for a numerical simulation must be obtained by
simulating the above described Kibble mechanism on a lattice. This
may be achieved by implementing a Monte-Carlo simulation
implemented first by Vachaspati and Vilenkin
\cite{Vachaspati:1984dz}. They considered a discretization of the
vacuum manifold ($S^1$) into three points and then assigned
randomly these three discrete phases to points on a square lattice
in physical space in three dimensions. For those square plaquettes
for which a complete winding of the phase in the vacuum manifold
occured, they assigned a string segment passing through. It may be
shown that this algorithm leads to strings with no ends within the
lattice volume \ie strings either form loops or go through the
entire lattice volume (infinite strings). This simulation showed
that the initial string network consists of 80\% long strings and
20\% loops.

In order to find the cosmological effects of this initial string
network, it must be evolved in time. Since it is impractical to
evolve the full field equations on the vast range of scales
$10^{-30} cm$ (string width) to $10^3 Mpc$ (largest cosmological
scales) we must resort to some approximation scheme. The ratio
$w/R$ of the width of the string over the string coherence scale
is an intrinsically small parameter for cosmological strings ($w/R
<< 10^{-30}$ for $t>t_{eq}$). Thus, this parameter can be used to
develop a perturbation
expansion\cite{f74,Arodz:1995dg,Gregory:1988qv} for the action
that describes the dynamics of a string segment. The zeroth order
approximation is obtained for $w/R = 0$. It is the generalization
of the point particle relativistic action to the string with mass
per unit length $\mu$ known as {\it Nambu action} and is defined
as: \be S=-\mu \int_A^B dl d\tau \sqrt{-g^{(2)}} + O(w/R) \ee
where $g^{(2)}= det(X_{,a}^\mu X_{,b}^\nu g_{\mu \nu})$  ($a,b =
1,2\rightarrow (\tau,l)$, $\mu,\nu=1,...4$) is the metric of the
world sheet spanned by the string ($X^\mu (\tau, l)$ is a
four-vector that determines the location of the string worldsheet
in 4d spacetime). The corrections to the  Nambu action are of
order of the string width $w$ divided by its curvature radius
(coherence length) $R$\cite{f74,Arodz:1995dg,Gregory:1988qv}. The
Nambu action is an excellent approximation to the dynamics of
non-intersecting cosmic string segments and is much simpler to
handle numerically than the full field theoretic action.

A crucial assumption in the derivation of the Nambu action is that
string segments do not interact with each other. This assumption
breaks down when two string segments intersect. The outcome of
such an event can only be found by evolving the full field
equations. Such numerical experiments have
shown\cite{Shellard:1987bv,m88} that at intersections string
segments {\it exchange parteners} (rather than passing through
each other) and a process known as {\it intercommuting} occurs.
Intercommuting is closely related to the right angle scattering of
vortices in head on collisions which has been observed in several
numerical
experiments\cite{Shellard:1987bv,m88,Samols:1991ne,Myers:1991yh,pinst}
and can be analytically understood using dynamics in moduli
spaces\cite{r88}. Intercommuting tends to favor the formation of
string loops which oscillate and decay by emitting gravitational
radiation with characteristic frequency $\omega \sim R^{-1}$ ($R$
is the loop radius) and rate ${\dot E} = \gamma G\mu^2$ ($\gamma =
const\simeq 50$)\cite{Vachaspati:1984gt,Caldwell:1991jj}. Once the
loop radius shrinks down to the string width (scale of symmetry
breaking) the loops decay to particles and may provide a mechanism
for explaining the observed ultra-high energy cosmic
rays\cite{Bhattacharjee:1990js}.

Therefore, this mechanism for loop formation, provides also an
efficient way for converting long string energy that redshifts
with the universe expansion as $E_{str} \sim a^{-2}$ to radiation
energy that redshifts as $E_{rad} \sim a^{-4}$. This is the
crucial feature that prohibits strings from dominating the energy
density of the universe during the radiation era and makes the
cosmic string theory a viable theory for structure formation in
the universe.

Using the Monte-Carlo initial conditions described above, the
Nambu action to evolve string segments and intercommuting to
describe intersection events, it is straightforward (though not
easy) to construct numerical simulations describing the evolution
of the string network in an expanding universe. Using this
approach it has been shown\cite{Albrecht:1984xv} that the initial
string network quickly relaxes in a robust way to a scale
invariant configuration known as {\it the scaling solution}.
According to the scaling solution the only scale that
characterizes the string network is the horizon scale at any given
time. On scales larger than the causal horizon scale $t$, the
network consists of a random walk of long strings which are
coherent on approximately horizon scales. On scales smaller than
the horizon more recent simulations
\cite{Allen:1990tv,Bennett:1987vf,Bennett:1989ak,Albrecht:1989mk}
have shown (see Fig. 1) that the network consists of a fixed
number of approximately 10 long strings coherent on horizon scales
and a large number of tiny loops with typical radius $10^{-4} t$.
The efficient formation of tiny loops by the intercommuting of
long strings on small scales leads to the existence of wiggles on
the long strings (wiggly strings) which further affect the
evolution of the network\cite{Vilenkin:1990mz}. The {\it
effective} mass per unit length of wiggly strings is larger than
the bare mass ($\mu_{eff} \simeq 1.4 \mu$) and their tension is
smaller than the bare tension\cite{Bennett:1989yp}.
\begin{figure}[h]
\centering
\includegraphics[bb=150 250 450 620,width=6.7cm,height=8cm,angle=-90]{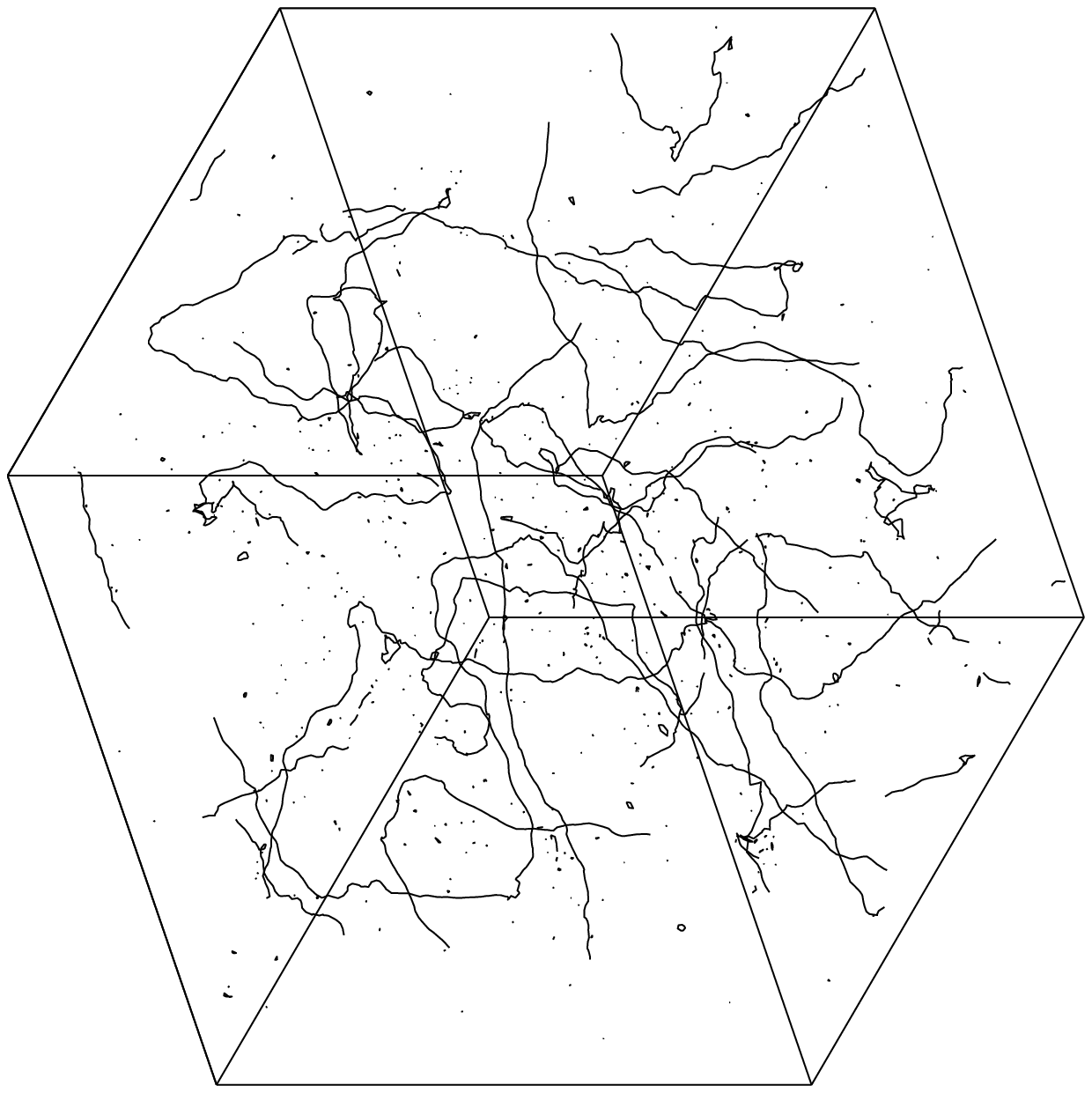}
\caption{The string network evolved during the matter era consists
of a dominant component of horizon sized long strings and a
distribution of fast tiny loops (from Ref. \cite{Allen:1990tv})}.
\label{fig1}
\end{figure}
There is a simple heuristic way to understand why does the network
of strings approach a scaling solution with a fixed number of long
strings per horizon scale. Consider for example the case when the
number of long strings per horizon increases drastically. This
will inevitably lead to more efficient intercommuting and loop
formation thus transfering energy from long strings to loops and
reducing the number of long strings per horizon back to its
equilibrium value. Similarly if the number of long strings
decreases reduced intercommuting and loop formation will tend to
increase the number of long strings towards an equilibrium value.
These heuristic arguments have been put in more detailed form
using differential equations in Ref. \cite{Vilenkin:1984ib}.
\subsection{Gravitational Effects: The Deficit Angle}
The most important interaction on cosmological scales is gravity.
It is therefore important to understand the gravitational effects
of strings in order to understand their cosmological predictions.
The straight string solution is thin, cylindrically symmetric and
Lorentz invariant for boosts along the length of the string. This
imposes the following constraint on components of the energy
momentum tensor $T_{\mu\nu}$ \be T_z^z (\rho) = T_0^0 (\rho)
\simeq \mu \delta (x) \delta (y) \ee Also \be T^\nu_{\mu,\nu} =0
\Rightarrow {d\over {dx}} T_x^x (\rho) = 0 \Rightarrow
T_x^x=T_y^y=0 \ee where use was made of the cylindrical symmetry
and of the fact that $T_x^x = T_y^y \rightarrow 0$ as
$r\rightarrow \infty$. Therefore, the string energy momentum
tensor may be approximated by \be T_{\mu \nu} \simeq \delta (x)
\delta (y) diag(\mu,0,0,-\mu)\equiv diag(\rho,p_x,p_y,p_z) \ee
which implies that the string has significant negative pressure
(tension) along the z direction \ie \be p_z = -\rho =-\mu
\hspace{1cm} p_x=p_y = 0 \ee This form of $T_{\mu\nu}$ may be used
to obtain the Newtonian limit for gravitational interactions of
strings with matter. For the Newtonian potential $\Psi$ we have
\be \nabla^2 \Psi = 4\pi G(\rho + \sum_i p_i)=0 \Rightarrow {\vec
F_N} = 0 \label{newtpot1}\ee and a test particle would feel no
force by a nearby motionless straight string.

Simulations have shown however that realistic strings are neither
straight nor motionless. Instead they have small scale wiggles and
move with typical velocities of $v_s\simeq 0.15c$ coherent on
horizon scales. What is the energy momentum tensor and metric of
such wiggly strings?

The main effect of wiggles on strings is to destroy Lorentz
invariance along the string axis, to reduce the effective tension
and to increase proportionaly the effective mass per unit length
of the string. Thus, the energy momentum tensor of a straight
wiggly string is \be T_{\mu \nu} \simeq \delta (x) \delta (y)
diag(\mu_{eff}, 0, 0, -T)\label{wigtmn} \ee with $T\equiv -p_z
<\mu$, $\mu_{eff} > \mu$ and $\mu_{eff} T = \mu^2$
\cite{Vachaspati:1991tt} ($\mu$ is the `bare' mass per unit length
obtained from the field Lagrangian). As expected the above $T_{\mu
\nu}$ reduces to the smooth straight string case for $\mu_{eff} =
T$. The breaking of Lorentz invariance by the wiggles also induces
a non-zero Newtonian force between the wiggly string at rest and a
test particle, since the tension (negative pressure) is not able
to completely cancel the effects of the energy density (Eq.
(\ref{newtpot1})). This may be seen more clearly by using the
Einstein's equations with the tensor of Eq. (\ref{wigtmn}) to find
the metric around a wiggly string. The result in the weak field
limit (small $G\mu$) is\cite{Vachaspati:1991tt} \be ds^2 =
(1+h_{00}) (dt^2 -dz^2 - (1-4G\mu_{eff})^2 r^2 d\varphi^2)
\label{strmet1}\ee with \be h_{00} = 4G(\mu_{eff} - T) ln(r/r_0)
\ee where $r_0$ is an integration constant. Clearly, in the
presence of wiggles ($\mu_{eff} \neq T$) the Newtonian potential
$h_{00}$ is non-zero. The change of the azimouthal variable
$\varphi$ to the new variable $\varphi' \equiv
(1-4G\mu_{eff})\varphi $ makes the metric (\ref{strmet1}) very
similar to the Minkowski metric with the crucial difference of the
presence of the Newtonian $h_{00}$ term and the fact that the new
azimouthal variable $\varphi'$ does not vary between $0$ and
$2\pi$ but between $0$ and $2\pi - 8\pi G\mu_{eff}$. Therefore
there is a {\it deficit angle} \cite{Vilenkin:1981zs} $\alpha =
8\pi G \mu_{eff}$ in the space around a wiggly or non-wiggly
string. Such a spacetime is called {\it conical} and leads to
several interesting cosmological effects especially for moving
long strings.

The main mechanism by which strings create perturbations that
could lead to large scale structure formation is based on velocity
perturbations created by moving long strings.
 Long, approximatelly straight strings moving with velocity
$v_s$ induce velocity perturbations to the surrounding matter
directed towards the surface swept in space by the string.

Using the geodesic equations in the spacetime of a moving long
strings it may be shown that the total velocity perturbation
induced by a moving long string to surrounding matter close to the
surface swept by the string  is \be \Delta v = {{2\pi G (\mu_{eff}
-T)}\over {\gamma_s v_s}} + 4\pi G \mu_{eff} v_s \gamma_s \ee
where the first term is due to the Newtonian interaction of the
string wiggles with matter while the second term is an outcome of
the conical nature of the spacetime.

Another particularly interesting effect of the string induced
deficit angle is the creation of a characteristic signature on CMB
photons. The type of this signature may be seen in a heuristic way
as follows:
 Consider a straight long
string moving with velocity $v_s$ between the surface of last
scattering occuring at the time of recombination $t_{rec}$ and an
observer at the present time $t_0$.  The presense of the moving
long string between the observer and the last scattering surface
induces an effective Doppler shift to the CMB photons due to the
deficit angle of the moving long string. For photons reaching the
observer through the `back' (`front') of the string we have \be
({{\delta T}\over T})_{1,(2)}  = \pm 4\pi G \mu_{eff} v_s \gamma_s
\ee where the $1 (2)$ and $+ (-)$ refer to photons passing through
the `back' (`front') of the string.

Therefore a moving long string present between $t_{rec}$ and today
induces line step-like discontinuities on the CMB sky with
magnitude\cite{Kaiser:1984iv,Gott:1984ef} \be {{T_1 - T_2} \over
T} = 8\pi G \mu_{eff} v_s \gamma_s \ee This effect is known as the
Kaiser-Stebbins effect. Notice that there is no Newtonian term
inversely proportional to the wiggly string velocity as was the
case for the induced velocity perturbations. As expected, photons
do not feel the Newtonian interaction with strings. This is a
crucial observation related to the formation of the Doppler peak
in the CMB spectrum by wiggly strings (see section 3).

Due to the conical nature of the spacetime around a straight
cosmic string it acts like a cylindrical gravitational
lens\cite{Vilenkin:1984ea} with a very unusual and characteristic
pattern of lensed images\cite{deLaix:1997jt}. The two images of a
source behind a string which creates a conical spacetime with
deficit angle $\delta=8\pi G\mu_{eff}$ are separated by an angle
\be \alpha = \frac{D_l}{D_s} \delta \sin \theta \ee where $D_s$ is
the angular diameter distance from us to the source, $D_l$ is the
distance between source and lens and $\theta$ is the angle between
the string tangent vector and the line of sight. These images are
predicted to appear at equal magnitudes with no magnification or
distortion (but at slightly different redshifts if the string is
moving) unlike the lensing patterns of ordinary gravitational
lenses. Therefore a gravitational lensing event induced by a
cosmic string is expected to involve a number of neighbouring
double images with typical separation of a few arcsec.

Having completed a brief review of the main cosmological effects
of cosmic strings I will now give a brief historical overview of
the development of the model. I first discuss in the next section
the development during the eighties: the {\it golden age} of the
cosmic string model. The strong constraints that practically rule
out the model obtained in the nineties will be discussed in
section 3 while recent lensing events that give some marginal hope
for the existence of lighter strings are discussed in section 4.
The conclusion of the review is given in section 5.

\section{THE RISE: MICROPHYSICS MEETS MACROPHYSICS}
The interest on cosmic strings as a possible theory for structure
formation originated in the early eighties with two papers by
Vilenkin \cite{Vilenkin:1981kz} and Zeldovich
\cite{Zeldovich:1980gh} who pointed out that loops of linear
defects produced during GUT phase transitions had the right
effective gravitational mass to act as seeds for the formation of
galaxies. This observation opened two major prospects:
\begin{itemize}
\item The potential verification of the GUT's via cosmological
observations \item The resolution of the long standing puzzle of
the origin the primordial fluctuations that gave rise to structure
formation
\end{itemize}
This was a prime example of a meeting point between macrophysics
(cosmology) and microphysics (particle physics). These prospects
gave the motivation for intense research activity on the subject
in the eighties. New papers appeared in the early eighties with
more encouraging results
\cite{Turok:1985tt,Turok:1984cn,Silk:1984xk,Vilenkin:1983jv}
showing that not only GUT string loops had the right mass to be
the seeds of galaxy formation but also they could produce the
observed galaxy-galaxy correlation function
\cite{Turok:1984cn,Turok:1985tt} and the observed cluster-cluster
correlation function. It was the first time that a model based on
concrete microphysical theories could make such successful
macrophysical predictions even if they were still qualitative at
that stage.  Cosmic strings could even resurrect neutrinos as dark
matter candidates (Hot Dark Matter (HDM)). Neutrinos were ruled
out as dark matter in the context of adiabatic perturbations
coming from inflation because they predicted that in contrast to
observations, smaller structures form by fragmentation of larger
ones\cite{bond84}. The problem was the free streaming of fast
moving neutrinos which wiped out galactic scale fluctuations.
Cosmic string seeds however could survive the free streaming of
neutrinos and form galaxies
earlier\cite{Brandenberger:1987er,Brandenberger:1987kf,Scherrer:1988id}
than larger structures in agreement with observations. Larger
planar structures due to long string wakes were also more
prominent with
HDM\cite{Perivolaropoulos:1989ug,Brandenberger:1989my} than with
CDM\cite{Stebbins:1987cy}.

These successes of the theory which mainly appeared in the first
half of the eighties decade lead to a dramatic increase in the
number of papers in the second half of the decade (see Fig. 2).
The number of papers with the word 'cosmic strings' in their title
from 23 in the first half of the decade (1980-1985) soared to 272
in the second half (1985-1990). This would be its peak for the
years to come.
\begin{figure}[h]
\centering
\includegraphics[bb=120 80 500 820,width=6.7cm,height=8cm,angle=-90]{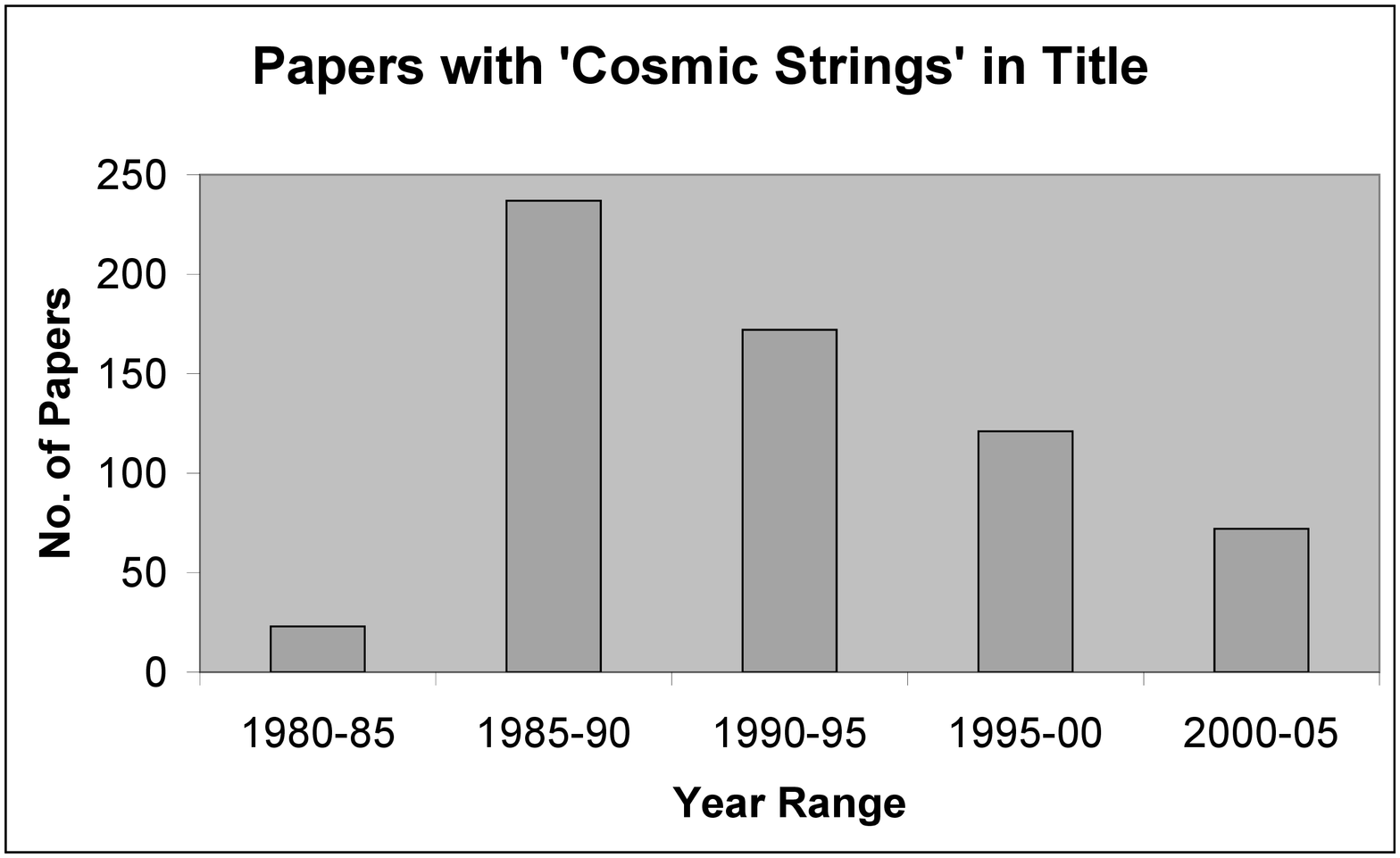}
\caption{The number of papers with the words 'cosmic strings' in
their title peaked during the second half of the eighties decade
(data from the 'spires'\cite{spires} database).} \label{fig2}
\end{figure}

It soon became clear to the community that the inherent
nonlinearities in the formation and cosmological evolution of
cosmic string networks make it difficult to make quantitative
predictions for detailed comparison with observations. In the mean
time, during the same decade, the competing model for structure
formation based on quantum fluctuations produced during inflation
developed much faster. The linear nature of inflationary
perturbations allowed detailed comparison with cosmological data
for galaxies and large scale structure. This comparison showed
that the scale invariant gaussian primordial perturbations
produced during inflation combined with $\Omega_0 = 1$ Cold Dark
Matter (CDM) were consistent with most cosmological observations
of the time and started to reduce the interest on cosmic strings
towards the end of the eighties\cite{Blumenthal:1984bp}.  The flat
gaussian CDM model based on inflation was established as the
Standard Model for structure formation. It became clear that
detailed and reliable numerical simulations of the cosmic string
network evolution were needed in order to keep the cosmic string
theory alive.

An early simulation of the cosmic string network evolution
\cite{Albrecht:1984xv} had indicated that at any time there was
one long string per horizon volume with curvature radius about the
horizon length and it would chop off by intercommuting about one
horizon scale loop per expansion time. This 'parent' loop would
then fragment into a few daughter loops which would live $10^4$
times longer before loosing all their energy to gravitational
radiation. As discussed in the introduction this energy loss
mechanism allows the network density to scale like the background
cosmological density and prevents it from dominating and
overclosing the universe. This type of network solution is known
as the {\it 'scaling solution'}. In this picture, the growth of
density inhomogeneities starts at the redshift of matter
domination $z_{eq}$ (assuming CDM) when most of the string density
is in slowly moving loops produced at earlier times. These loops
were shown to have a number density and correlation function
comparable to that of bright
galaxies\cite{Turok:1984cn,Turok:1985tt}. Since most of the string
mass was found to be in loops, the accretion around the much less
dense long strings was ignored.

Even though this 'one loop - one object' picture produced a mass
function of haloes that fitted both the number density of bright
galaxies and the number density of clusters with the same value of
mass per unit length $\mu$\cite{Turok:1985tt} predicted by GUT's
it was realized to be incorrect towards the end of the eighties by
detailed numerical simulations of string network evolution. These
simulations were performed independently by three groups
\cite{Allen:1990tv,Bennett:1987vf,Bennett:1989ak,Albrecht:1989mk,Bennett:1989yp}
and showed that the efficiency with which a network of long
strings chops off loops is much less than previously thought. A
much denser network of long strings is thus required to chop off
enough decaying loops to keep the long strings scaling with the
cosmological density. In addition it was found that the size of
the produced loops at any given time is much smaller and their
number density is much greater. These results changed drastically
the original 'one loop-one object' picture described above. The
small and fast loops produced after the time of equal matter and
radiation $t_{eq}$ would produce linear wakes at least tens of
$Mpc$ long. It is implausible to associate such an elongated
structure to a single galaxy. Thus it was realized that the
dominant perturbations produced by the cosmic string network are
due to horizon sized long strings which produce velocity
perturbations that lead to planar overdensities (wakes) as they
move through the homogeneous cosmological background. According to
the new simulations there are about ten horizon sized long strings
per horizon volume at any given time. Due to the production of
several small loops these strings are not smooth but have a wiggly
structure (this picture however has been disputed again by more
recent simulations\cite{Hindmarsh:1997qy} indicating that strings
may indeed be smooth!). The planar wakes produced by these strings
give rise to a network of planar overdensities with a
distinguished scale of about $40\times 40 Mpc^2$ and thickness of
about $4Mpc$\cite{Stebbins:1987cy} which were more pronounced in
cosmic string models with HDM \cite{Perivolaropoulos:1989ug}.

\section{THE FALL: OBSERVATIONAL CONSTRAINTS}

The Kaiser-Stebbins temperature discontinuities produced in the
CMB sky by long cosmic strings are thought to be the
characteristic imprint of a cosmic string network on the CMB. The
first attempt towards obtaining a detailed pattern of these
discontinuities was made in Ref. \cite{Bouchet:1988hh} where the
evolution of CMB photons was simulated through an evolving string
network from $t_{rec}$ to the present time $t_0$. This type of
simulations was further improved in the nineties
\cite{Bennett:1992ed,Contaldi:1998mx,Allen:1997ag} and the full
cosmic string CMB spectrum was derived showing no evidence for
acoustic oscillations (Doppler peak). However, all such numerical
approaches suffer from an important drawback: Their dynamic range
can not extend as to cover the required cosmic expansion factor of
$10^4$ due to the vast range of scales involved. This problem is
verified with the latest such string simulations which show a
dependence of the string CMB normalization on the resolution of
the simulation used\cite{Landriau:2003xf}. Thus, some type of
network modeling is required in order to obtain detailed
observational predictions including small scales. There have been
several attempts to construct such analytical network
models\cite{analyt-evol,Perivolaropoulos:1992if,Austin:1993rg}
capturing the main features of the new string evolution
simulations. Early encouraging analytical results using this type
of models indicated that on large scales cosmic strings predict a
scale invariant
spectrum\cite{Perivolaropoulos:1992if,Perivolaropoulos:1994ry,Gilbert:1995de}
(which is in agreement\cite{Perivolaropoulos:1992if} with the COBE
spectrum) with well defined non-gaussian
features\cite{Perivolaropoulos:1992gy,Perivolaropoulos:1993vu,Gangui:1994wh,Moessner:1993za}.
In addition to deriving the predicted CMB scale invariant spectrum
on large scales these early studies showed that wiggly gauged
strings predict the existence of a Doppler
peak\cite{Perivolaropoulos:1994ry} in the CMB spectrum in contrast
to global defects\cite{Durrer:1998rw} where decoherence dominates
and leads to a smearing of the Doppler peak. The origin of this
peak is the wiggly nature of long strings whose Newtonian
interaction affects the plasma more significantly than it affects
the photons between $t_{rec}$ and the present
time\cite{Perivolaropoulos:1994ry}. The existence of such a peak
was confirmed by more recent studies\cite{Pogosian:1999np} where
analytical models were fed as input into standard cosmological
evolution codes for the detailed derivation of the full CMB and
LSS power spectra from primordial fluctuations. It was found that
the Doppler peak is lower compared to observations and shifted
towards smaller scales (larger $l$). The coherence scale of long
strings is significantly smaller than the horizon at $t_{rec}$.
The derived spectra were then
compared\cite{Pogosian:1999np,Pogosian:2003mz} with the
corresponding observational data. As discussed below these results
constrain the string contribution to the primordial fluctuations
to be less than $10\%$.

This severe constraint on the cosmic string contribution to the
primordial fluctuations can be translated into corresponding
constraints onto the models that predict cosmic string formation
after hybrid inflation\cite{Copeland:1994vg,Tkachev:1998dc}. This
prospect gains importance in view of the fact that string theory
or M-theory
demands\cite{Rocher:2004my,Jeannerot:2003qv,Dvali:2003zj,Sarangi:2002yt}
the formation of a network of cosmic strings after a period of
inflation induced by the collision of branes that merge to form
the brane on which we live. Thus, according to these models the
CMB temperature power spectrum emerges as a superposition of
adiabatic gaussian perturbations coming from brane inflation and
cosmic string perturbations with weighting factors $W$ and $B$
respectively ie \be C_l=W C_l^{adiabatic} + B C_l^{strings} \ee
Comparison of the predicted forms of $C_l^{adiabatic}$ and
$C_l^{strings}$ with the WMAP observed spectrum (Fig. 3) shows
that in a $\Lambda CDM$ universe, the spectrum is well fit by pure
adiabatic perturbations\cite{Jaffe:2000tx,Spergel:2003cb} and
$B=0$ at the $90\%$ confidence level \cite{Pogosian:2003mz}. A
tiny contribution from strings (about $5\%$ can slightly improve
the fit to the data\cite{Contaldi:1998qs} but it is constrained to
$B\leq 0.09$ at the $99\%$ confidence level
\cite{Pogosian:2003mz,Pogosian:2004ny,Wu:2005ap}. The constraint
on $B$ translates to a constraint on the effective string mass
parameter $\mu$ as \be G\mu \leq 1.3 \times 10^{-6} \sqrt{B
\lambda/0.1} \ee where $\lambda$ is a dimensionless measure of the
intercommutation rate (the probability of string segments
exchanging partners at colision) which is less than 1 in models
with extra dimensions like brane inflation. The constraint on the
parameter $G\mu$ (the single parameter of the cosmic string model)
constrains also the magnitude of the predicted Large Scale
Structure (LSS) power
spectrum\cite{Pogosian:2003mz,Albrecht:1997nt} to less that $10\%$
of the observed one (Fig. 4). The LSS spectrum constraints however
can be made weaker in an open universe\cite{Avelino:1997hy}
according to a study based of numerical evolution of the string
network in an open CDM cosmology.

\begin{figure}[h]
\centering
\includegraphics[bb=90 200 550 680,width=6.7cm,height=8cm,angle=0]{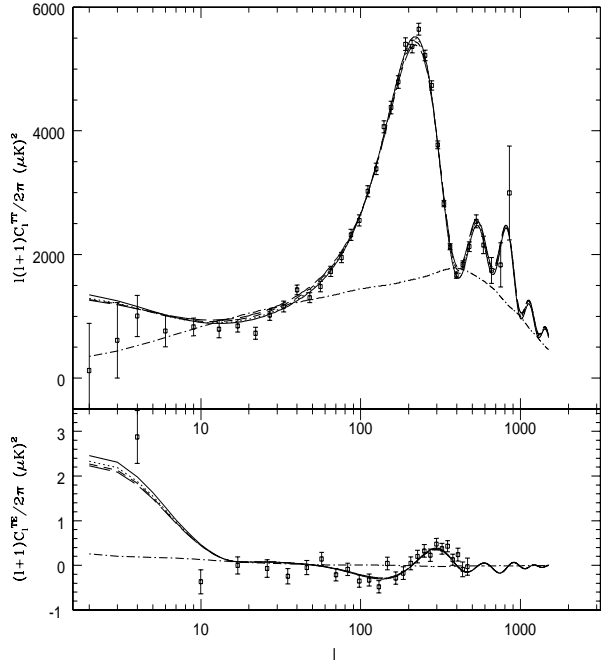}
\caption{The CMB power spectrum ($C_l^{TT}$ upper and $C_l^{TE}$
lower) predicted in the cosmic string theory (dot-dashed lines) is
significantly different compared to the spectrum observed by WMAP
which practically coincides with the inflationary predictions
(from Ref. \cite{Pogosian:2003mz}).} \label{fig3}
\end{figure}

Comparable or even stronger constraints on $G\mu$ can be imposed
by a direct search for the predicted Kaiser-Stebbins temperature
discontinuities predicted to exist in CMB maps. Despite
preliminary encouraging results\cite{Perivolaropoulos:1998fu}
utilizing the COBE data, a recent detailed statistical
analysis\cite{Jeong:2004ut} searching for such signatures in the
WMAP high resolution CMB maps found no such non-gaussian signal.
This negative result is translated to a stringent constraint of
$G\mu$ \be G\mu \leq 3.3 \times 10^{-7} \label{smb}\ee

\begin{figure}[h]
\centering
\includegraphics[bb=90 200 550 680,width=6.7cm,height=8cm,angle=0]{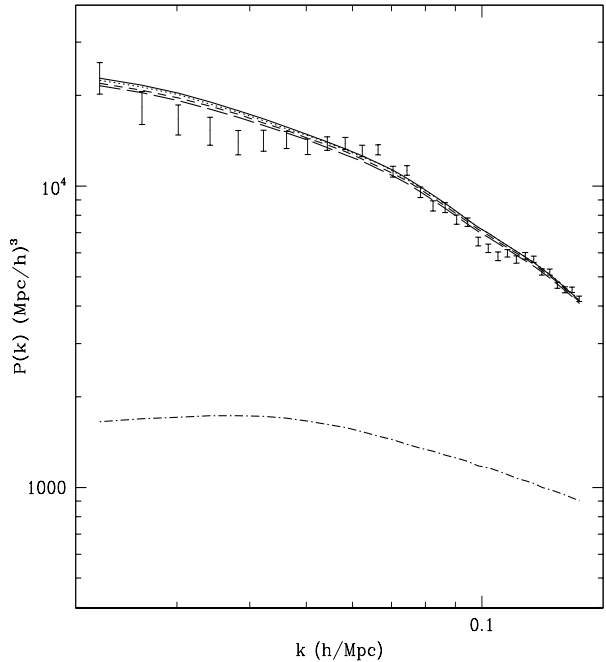}
\caption{The cosmic string predicted Large Scale Structure power
spectrum normalized to the large CMB scales (dot-dashed line) does
not have enough power to mach observations which are well fitted
by inflationary predictions (from Ref. \cite{Pogosian:2003mz}).}
\label{fig4}
\end{figure}

Accelerating massive string segments are expected to produce
gravitational waves leading to a stochastic gravitational wave
background produced by the string network. Such a background would
distort the regularity of the pulsar timing which is not observed.
This lack of distortion puts an upper limit to the energy density
of stochastic gravitational waves which is translated to a tight
bound \cite{Vachaspati:1984gt,Lommen:2002je} on $G\mu$ as \be G\mu
\leq 10^{-7} \label{gwb}\ee There are uncertainties involved in
this bound however because its main source is the loop component
and the small scales of the string network which is the most
uncertain part of the sting network simulations.

These constraints on the string network are easily translated into
useful constraints for the brane inflation models that predict its
formation. To rescue brane inflation various mechanisms  have been
proposed  which either consider more complicated models, or
require additional ingredients so that cosmic strings are not
produced at the end of inflation \cite{Urrestilla:2004eh}. More
optimistic approaches have claimed that even without
modifications, the cosmic string contribution is subdominant
anyway for small enough couplings and is therefore consistent with
the data\cite{Rocher:2004my}.

\section{THE HOPE: GRAVITATIONAL LENSING EVENTS}
Despite the above severe constraints which have significantly
decreased the interest in the cosmic string theory as a model
contributing to structure formation, the possibility of lighter
strings existing in the universe remains and can not be excluded.
There are two observational sources that can lead to detection of
such lighter strings: gravitational lensing and gravitational
radiation.

Lensing events consistent with a cosmic string playing the role of
the lense had been reported a few years ago \cite{Cowie87} but a
more careful analysis indicated that they were not due to a
string. More recent searches \cite{Shirasaki:2003ni} for string
induced lensed pairs have lead to negative results. However a new
candidate event named CSL-1 (Capodimonte-Sternberg Lens candidate
no.1) was reported a couple of years ago by a Russian-Italian
collaboration \cite{Sazhin:2003cp,Sazhin:2004fv}. The most
intriguing property of CSL-1 (Fig. 5) is that the object is
clearly extended and the isophotes of the two sources show no
distortion at all. Also both images have a redshift of 0.46 and
the two spectra are identical at a $99.96\%$ confidence level.
Extensive modeling of the photometric properties of CSL-1 showed
that no traditional lensing by compact lens can satisfactorily fit
the observed morphology. The most likely explanation of the
observed properties of CSL-1 is that it is the result of lensing
by an interposed cosmic string. Such a string should have \be G\mu
\geq 4\times 10^{-7}\ee to produce the observed image separation
($2''$) at the observed redshift $z=0.46\pm 0.008$. This range is
marginally in conflict with the statistical CMB bounds coming from
the lack of linear discontinuities in the WMAP maps (see eq
(\ref{smb})) and with the corresponding bound coming from the
stochastic gravitational wave background of eq (\ref{gwb}). In a
more recent publication\cite{Sazhin:2004fv} the CSL collaboration
reported a significant excess of lensed pairs in the field of
CSL-1 ($16' \times 16'$ spaned by $4000\times 4000$ pixels) over
the no more than two lensed pairs expected due to conventinal
lensing objects such as galaxies.
\begin{figure}[h]
\centering
\includegraphics[bb=85 60 300 140,width=4cm,height=2cm,angle=0]{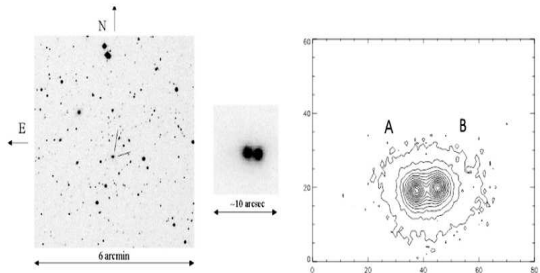}
\caption{The CSL-1 lensing event may indicate the presence of a
string lens because the two sources are undistorted as indicated
by the isophotes and have identical spectra.} \label{fig5}
\end{figure}
In the case of a straight string approximatelly 9 lensed pairs
would be expected in the field, going up to 200 pairs for a random
walk shape of the string. The observed number of pairs was 11 but
they were not concentrated on a roughly linear strip. A detailed
spectroscopic analysis (in progress by the CSL collaboration) is
required in order to resolve the issue.

\section{CONCLUSION-OUTLOOK}
Scientific progress is achieved by continous interaction between
theory and experiment or observation. Theoretical models make
predictions which are tested by experiments or observation thus
confirming or ruling out the corresponding models. In order to
achieve such progress the models should be defined well enough to
make unique predictions and the experiment should be accurate
enough to determine if these predictions are realized in nature.
Until a couple of decades ago neither of these conditions were
fulfilled for cosmology. Models had too many parameters to make
unique predictions and observations were not accurate enough to
conclusively test the theoretical predictions. For example scalar
field inflation has in principle an infinite number of parameters
since the dynamics of the scalar field is determined by an
unspecified potential. Thus even though generically inflation
predicts a flat space with scale invariant gaussian primordial
perturbations it is easy to chose parameters leading to an open
universe with non-gaussian perturbations and a tilted spectrum.
Even the most precise observations would have a hard time to rule
out such a model.

In contrast to inflation, the cosmic string theory for
cosmological perturbations has a single free parameter $G\mu$
whose value is well constrained by microphysics to $G\mu \simeq
10^{-6}$. This is a well defined theory whose predictions were
subject to detailed comparison with accurate cosmological
observations during the past decade. From this comparison it
became clear that the model is ruled out and at best it could have
a tiny contribution (less than $9\%$) to the primordial
perturbation spectrum. Such a healthy example of scientific
progress through the interaction between theory and observations
is unfortunately not very frequent even during the present era of
precision cosmology. The cosmic string theory had the luck to
consist one of the few exceptions to this rule.

\end{document}